# Approximate $l$-State Solution of the Rotating Trigonometric Pöschl-Teller Potential

Majid Hamzavi[1*], Sameer M. Ikhdair[2**]

[1]*Department of Basic Sciences, Shahrood Branch, Islamic Azad University, Shahrood, Iran*

[2]*Physics Department, Near East University, 922022 Nicosia, North Cyprus, Mersin 10, Turkey*

\* *majid.hamzavi@gmail.com*

\*\* Corresponding author *sikhdair@neu.edu.tr*

**Abstract**

The trigonometric Pöschl-Teller (PT) potential describes the diatomic molecular vibration. We have obtained the approximate solutions of the radial Schrödinger equation (SE) for the rotating trigonometric PT potential using the Nikiforov-Uvarov (NU) method. The energy eigenvalues and their corresponding eigenfunctions are calculated for arbitrary $l$-states in closed form. In the low screening region, when the screening parameter $\alpha \to 0$, the potential reduces to Kratzer potential. Further, some numerical results are presented for several diatomic molecules.

## 1. Introduction

The solution of the fundamental dynamical equations is an interesting phenomenon in many fields of physics and chemistry. The exact solutions of the SE for a hydrogen atom (Coulombic) and for a harmonic oscillator represent two typical examples in quantum mechanics [1-3]. The Mie-type and pseudoharmonic potentials are also two exactly solvable potentials [4-5]. Many authors have exactly solved SE with different potentials and methods [6-16].

The trigonometric PT potential proposed for the first time by Pöschl and Teller [17] in 1933 was to describe the diatomic molecular vibration. Chen [18] and Zhang et al. [19] have studied the relativistic bound state solutions for the trigonometric PT potential and hyperbolical PT (Second PT) potential, respectively. Liu et al. [20] studied the trigonometric PT potential within the framework of the Dirac theory. Very recently, Hamzavi and Rajabi studied the exact $s$-wave solution $(l=0)$ of the Schrödinger equation for the vibrational trigonometric PT potential [21]. This potential takes the following form:

$$V(r) = \frac{V_1}{\sin^2(\alpha r)} + \frac{V_2}{\cos^2(\alpha r)}, \quad V_1 > 0, \quad V_2 > 0 \tag{1}$$

where the parameters $V_1$ and $V_2$ describe the property of the potential well while the parameter $\alpha$ is related to the range of this potential [20]. We find out that this potential has a minimum value at $r_0 = \frac{1}{\alpha}\tan^{-1}\left(\sqrt[4]{\frac{V_1}{V_2}}\right)$. When $V_1 = V_2$ then minimum value becomes at $r_0 = \frac{\pi}{4\alpha} \in (0, \infty)$ for $\alpha > 0$. The second derivative which determines the force constants at $r = r_0$ is given by

$$\left.\frac{d^2V}{dr^2}\right|_{r=r_0} = \frac{8\alpha^2\left(V_2 + \sqrt{V_1V_2}\right)}{\cos^2\left[\tan^{-1}\left(\sqrt[4]{\frac{V_1}{V_2}}\right)\right]}, \tag{2}$$

for any $\alpha$ value and

$$V(r_0) = \frac{\sqrt{V_1V_2} + V_2}{\cos^2\left[\tan^{-1}\left(\sqrt[4]{\frac{V_1}{V_2}}\right)\right]}, \tag{3}$$

which means that $V(r)$ at $r = r_0$ has a relative minimum for $\alpha > 0$. When $V_1 = V_2 = V$ then minimum value is $V(r_0) = 4V$ and $\left.\frac{d^2V}{dr^2}\right|_{r=r_0} = 32\alpha^2V$. In Figure 1, we draw the trigonometric PT potential (1) for parameter values $V_1 = 5.0\,fm^{-1}$, $V_2 = 3.0\,fm^{-1}$, $\alpha = 0.02\,fm^{-1}$. Here the potential has a minimum value at $r_0 = 0.27027\pi/\alpha$. The curve is nodeless in $\alpha r \in (0, \pi/2)$. For example, with $\alpha = 0.30\,fm^{-1}$, $r_0 = 2.8303\,fm$ and minimum potential $V(r_0 = 2.8303\,fm) = 15.746\,fm^{-1}$. It is worthy to note that in the limiting case when $\alpha \to 0$, the trigonometric PT potential can be reduced to the Kratzer potential [21,22]

$$V(r) = D_e \left(\frac{r - r_e}{r}\right)^2 + \eta,$$

where $r_e$ is the equilibrium intermolecular separation and $D_e$ is the dissociation energy between diatomic molecules. In our case, $D_e = V_1$, $\eta = V_2$ and $r_e = 1/\alpha$. In the case of $\eta = 0$ reduces to the molecular potential which is called the modified Kratzer potential proposed by Simons *et al.* [23] and Molski and Konarski [24]. In the case of $\eta = -D_e$, this potential turns into the Kratzer potential, which includes an attractive Coulomb potential and a repulsive inverse square potential, introduced by Kratzer in 1920 [25].

The aim of the present work is to extend our previous work [26] to the case of $l \neq 0$ (rotational case). We introduce a convenient approximation scheme to deal with the strong singular centrifugal term. The ansätz of this approximation possesses the same form of the potential and is singular as the centrifugal term $r^{-2}$. Thus, the Schrödinger equation with the trigonometric PT potential is solved approximately for its energy eigenvalues and corresponding wave functions with arbitrary rotation-vibration $(n,l)$ state [27].

This work is arranged as follows: in Section 2, the NU method with all the necessary formulae used in the calculations is briefly introduced and a shortcut of the method is given in Appendix A. In Section 3 we solve the SE for the given trigonometric PT potential and obtain its energy eigenvalues and the corresponding wave functions. Some numerical results are obtained for any arbitrary vibration-rotation quantum numbers $n$ and $l$. Finally, the relevant conclusion is given in Section 4.

## 2. NU method

The NU method can be used to solve second order differential equations with an appropriate coordinate transformation $s = s(r)$ [28]

$$\psi_n''(s) + \frac{\tilde{\tau}(s)}{\sigma(s)}\psi_n'(s) + \frac{\tilde{\sigma}(s)}{\sigma^2(s)}\psi_n(s) = 0 \tag{4}$$

where $\sigma(s)$ and $\tilde{\sigma}(s)$ are polynomials, at most of second-degree, and $\tilde{\tau}(s)$ is a first-degree polynomial. To find a particular solution of Eq. (4), using the separation of variables, one can insert the transformation $\psi_n(s) = \phi(s)y_n(s)$ to reduces the above equation into the form of the following hypergeometric type

$$\sigma(s)y_n''(s) + \tau(s)y_n'(s) + \lambda y_n(s) = 0 \tag{5}$$

Furthermore, the function $\phi(s)$ is defined by the logarithmic derivative

$$\frac{\phi'(s)}{\phi(s)} = \frac{\pi(s)}{\sigma(s)} \tag{6}$$

And the second part function $y_n(s)$ is in the form of a hypergeometric-type function whose solutions are given by Rodrigues relation

$$y_n(s) = \frac{B_n}{\rho(s)}\frac{d^n}{ds^n}[\sigma^n(s)\rho(s)] \tag{7}$$

where $B_n$ is the normalization constant and $\rho(s)$ is the weight function that satisfies the condition [28]

$$\frac{d}{ds}w(s) = \frac{\tau(s)}{\sigma(s)}w(s), \qquad w(s) = \sigma(s)\rho(s) \tag{8}$$

The function $\pi(s)$ and the parameter $\lambda$, required for this method, are defined as follows

$$\pi(s) = \frac{\sigma' - \tilde{\tau}}{2} \pm \sqrt{\left(\frac{\sigma' - \tilde{\tau}}{2}\right)^2 - \tilde{\sigma} + k\sigma} \tag{9a}$$

$$\lambda = k + \pi'(s) \tag{9b}$$

In order to find the value of $k$, the expression under the square root must be square of polynomial. Thus, a new eigenvalue equation is

$$\lambda = \lambda_n = -n\tau' - \frac{n(n-1)}{2}\sigma'' \tag{10}$$

where

$$\tau(s) = \tilde{\tau}(s) + 2\pi(s) \tag{11}$$

and its derivative must be negative [28]. In this regard, one can also derive the parametric generalization version of the NU method [29] as displayed in Appendix A.

## 3. The solution of radial SE for the trigonometric Pöschl-Teller potential

To study any quantum physical model characterized by the empirical molecular potential given in Eq. (1), we need to solve the following SE given by [1-2]

$$\left(\frac{P^2}{2m} + V(r)\right)\psi_{n,l,m}(r,\theta,\varphi) = E_{nl}\psi_{n,l,m}(r,\theta,\varphi), \tag{12}$$

where the potential $V(r)$ is taken as the trigonometric PT potential (1). Using the separation of variables by applying the wave function $\psi(r,\theta,\varphi) = \frac{1}{r}R_{n,l}(r)Y_{lm}(\theta,\varphi)$, we obtain the radial SE as

$$\left[\frac{d^2}{dr^2} + \frac{2m}{\hbar^2}\left(E_{nl} - \frac{V_1}{\sin^2(\alpha r)} - \frac{V_2}{\cos^2(\alpha r)}\right) - \frac{l(l+1)}{r^2}\right]R_{n,l}(r) = 0, \; r > 0 \tag{13}$$

where the radial wave function $R_{n,l}(r)$ has to satisfy the required boundary conditions, namely, $R_{n,l}(0) = 0$ and $R_{n,l}(\pi/2) = 0$ on the edges. Since the SE with the trigonometric PT potential has no analytical solution for $l \neq 0$ states, we resort to use an appropriate approximation scheme to deal with the centrifugal potential term as

$$\frac{1}{r^2} = \lim_{\alpha \to 0} \alpha^2 \left( d_0 + \frac{1}{\sin^2(\alpha r)} \right), \quad 0 < \alpha r < \pi/2 \tag{14}$$

where $d_0 = 1/12$ is a dimensionless shifting parameter and $\alpha r \ll 1$. The approximation (14) is done on the basis that $\sin(z) = z - z^3/3! + z^5/5! - z^7/7! + \cdots$, and in the limit when $z \to 0$, $\sin(z) \approx z$. To show the validity and accuracy of our choice to the approximation scheme (14), we plot the centrifugal potential term $1/r^2$ and its approximations: $\alpha^2/\sin^2(\alpha r)$ and $\alpha^2 \left( d_0 + 1/\sin^2(\alpha r) \right)$ in Figure 2. As illustrated, the three curves coincide together and show how accurate is this replacement. One of us has treated this problem in his recent work (see Ref. [30]). The insertion of the approximation (14) in Eq. (13) gives

$$\left[ \frac{d^2}{dr^2} + \varepsilon_{nl} - \frac{V_1'}{\sin^2(\alpha r)} - \frac{V_2'}{\cos^2(\alpha r)} - l(l+1)\alpha^2 \left( d_0 + \frac{1}{\sin^2(\alpha r)} \right) \right] R_{n,l} = 0, \tag{15a}$$

$$\varepsilon = \frac{2mE_{nl}}{\hbar^2}, \quad V_1' = \frac{2mV_1}{\hbar^2} \text{ and } V_2' = \frac{2mV_2}{\hbar^2}. \tag{15b}$$

To solve Eq. (15a) via the NU method, we need to change the variables as $s = \sin^2(\alpha r)$ to rewrite Eq. (15a) in a more convenient form amendable to NU solution:

$$\frac{d^2 R_{n,l}(s)}{ds^2} + \frac{\frac{1}{2} - s}{s(1-s)} \frac{dR_{n,l}(s)}{ds} + \frac{1}{s^2(1-s)^2} \left[ -As^2 + Bs - C \right] R_{n,l}(s) = 0, \tag{16}$$

where $R_{n,l}(s) \simeq R_{n,l}(r)$ and also we have defined

$$A = \frac{1}{4\alpha^2}\left(\varepsilon_{nl} - l(l+1)\alpha^2 d_0\right), \qquad (17a)$$

$$B = \frac{1}{4\alpha^2}\left(\varepsilon + V_1' - V_2' + l(l+1)\alpha^2(1-d_0)\right), \qquad (17b)$$

$$C = \frac{1}{4\alpha^2}\left(V_1' + l(l+1)\alpha^2\right). \qquad (17c)$$

Comparing Eq. (16) and relation (A2), we can easily obtain the coefficients $c_i$ ($i = 1, 2, 3$) as follows

$$c_1 = \frac{1}{2},\ c_2 = 1,\ c_3 = 1. \qquad (18a)$$

The values of the remaining coefficients $c_i$ ($i = 4, 5, ..., 13$) are found from the relation (A5) of Appendix A. In addition, the specific values of the coefficients $c_i$ ($i = 1, 2, ..., 13$) are listed as

$$c_4 = \frac{1}{4},\ c_5 = -\frac{1}{2},\ c_6 = \frac{1}{4}\left[1 + \frac{\varepsilon}{\alpha^2} - l(l+1)d_0\right],$$

$$c_7 = -\frac{1}{4}\left[(2l+1)^2 + \frac{1}{\alpha^2}\left(\varepsilon + V_1' - V_2'\right) - l(l+1)d_0\right],\ c_8 = \frac{1}{16}\left[(2l+1)^2 + \frac{4V_1'}{\alpha^2}\right],$$

$$c_9 = \frac{1}{16}\left(1 + \frac{4V_2'}{\alpha^2}\right),\ c_{10} = \frac{1}{2}\sqrt{(2l+1)^2 + \frac{4V_1'}{\alpha^2}},\ c_{11} = \frac{1}{2}\sqrt{1 + \frac{4V_2'}{\alpha^2}},$$

$$c_{12} = \frac{1}{4}\left[1 + \sqrt{(2l+1)^2 + \frac{4V_1'}{\alpha^2}}\right]\ \text{and}\ c_{13} = \frac{1}{4}\left(1 + \sqrt{1 + \frac{4V_2'}{\alpha^2}}\right). \qquad (18b)$$

By using the relation (A10), we can obtain the energy eigenvalues of the rotating trigonometric PT potential as

$$E_{nl} = \frac{\hbar^2\alpha^2 l(l+1)d_0}{2m} + \frac{2\hbar^2\alpha^2}{m}\left[n + \frac{1}{2} + \frac{1}{4}\left(\sqrt{(2l+1)^2 + \frac{8mV_1}{\hbar^2\alpha^2}} + \sqrt{1 + \frac{8mV_2}{\hbar^2\alpha^2}}\right)\right]^2. \qquad (19)$$

In the limit when $\alpha \to 0$, the energy formula (19) reduces into a constant value:

$$\lim_{\alpha \to 0} E_{nl} = \left(\sqrt{V_1} + \sqrt{V_2}\right)^2. \tag{20}$$

Further, there is no less of generality if $d_0 = 0$, then Eq. (19) becomes

$$E_{nl} = \frac{2\hbar^2 \alpha^2}{m}\left[n + \frac{1}{2} + \frac{1}{4}\left(\sqrt{(2l+1)^2 + \frac{8mV_1}{\hbar^2 \alpha^2}} + \sqrt{1 + \frac{8mV_2}{\hbar^2 \alpha^2}}\right)\right]^2, \tag{21}$$

where $n = 0, 1, 2, \cdots$ and $l = 0, 1, 2, \cdots$ are the vibration and rotation quantum numbers, respectively. To obtain a numerical energy eigenvalues for the present potential model, we take the following set of parameter values; namely, $m = 10 \, fm^{-1}$, $V_1 = 5.0 \, fm^{-1}$, $V_2 = 3.0 \, fm^{-1}$ and $\alpha = 1.2, 0.8, 0.4, 0.2, 0.02, 0.002$ [20]. As seen from Table 1, in the limit when potential range parameter $\alpha$ approaches zero, the energy eigenvalues approaches a constant value given by Eq. (20). We take $d_0 = 0$ and $d_0 = 1/12$, respectively. In Figure 3, we show the variation of the lowest vibration-rotation $1s$, $1p$, $2s$, $2p$, $3s$ and $3p$ states with the screening parameter $\alpha$ for a set of parameter values $m = 10 \, fm^{-1}$, $V_1 = 5.0 \, fm^{-1}$, $V_2 = 3.0 \, fm^{-1}$ and $d_0 = 0$. Further, for the same set of parameters, we draw the energy states versus the vibration quantum number $n$ in Figure 4.

Next, we need to calculate the wave functions. Using Eq. (18b) together with the relations (A11) and (A12) of Appendix A, we obtain the functions

$$\rho(s) = s^{\frac{1}{2}\sqrt{(2l+1)^2 + \frac{8mV_1}{\hbar^2 \alpha^2}}} (1-s)^{\frac{1}{2}\sqrt{1 + \frac{8mV_2}{\hbar^2 \alpha^2}}}, \tag{22}$$

$$\phi(s) = s^{\frac{1}{4}\left[1 + \sqrt{(2l+1)^2 + \frac{8mV_1}{\hbar^2 \alpha^2}}\right]} (1-s)^{\frac{1}{4}\left(1 + \sqrt{1 + \frac{8mV_2}{\hbar^2 \alpha^2}}\right)}. \tag{23}$$

Further, the relation (A13) gives the first part of the desired wave function:

$$y_n(s) = P_n^{\left(\frac{1}{2}\sqrt{(2l+1)^2 + \frac{8mV_1}{\hbar^2 \alpha^2}}, \frac{1}{2}\sqrt{1 + \frac{8mV_2}{\hbar^2 \alpha^2}}\right)} (1 - 2s), \tag{24}$$

and employing $R_{n,l}(s) = \phi(s) y_n(s)$, we finally get the radial wave functions from the relation (A14) as

$$R_{n,l}(s) = s^{\frac{1}{4}\left[1+\sqrt{(2l+1)^2 + \frac{8mV_1}{\hbar^2\alpha^2}}\right]} (1-s)^{\frac{1}{4}\left(1+\sqrt{1+\frac{8mV_2}{\hbar^2\alpha^2}}\right)} P_n^{\left(\frac{1}{2}\sqrt{(2l+1)^2+\frac{8mV_1}{\hbar^2\alpha^2}},\frac{1}{2}\sqrt{1+\frac{8mV_2}{\hbar^2\alpha^2}}\right)}(1-2s). \qquad (25)$$

Inserting $s = \sin^2(\alpha r)$ in the above equation, we get

$$R_{n,l}(r) = N_{nl}(\sin(\alpha r))^{(1+\eta_l)/2} (\cos(\alpha r))^{(1+\delta)/2} P_n^{(\eta_l/2,\delta/2)}(\cos(2\alpha r)), \qquad (26a)$$

$$\eta_l = \sqrt{(2l+1)^2 + \frac{8mV_1}{\hbar^2\alpha^2}}, \quad \delta = \sqrt{1+\frac{8mV_2}{\hbar^2\alpha^2}} \qquad (26b)$$

where $N_{nl}$ is a normalization factor to be calculated from the normalization conditions. For example, the ground $s$-state has the wave function:

$$R_{0,0}(r) = N_{00}(\sin(\alpha r))^{(1+\eta_0)/2} (\cos(\alpha r))^{(1+\delta)/2} P_0^{(\eta_0/2,\delta/2)}(\cos(2\alpha r)), \qquad (27a)$$

where $\eta_0 = \sqrt{1 + \frac{8mV_1}{\hbar^2\alpha^2}}$,

and for $1p$-state:

$$R_{0,1}(r) = N_{01}(\sin(\alpha r))^{(1+\eta_1)/2} (\cos(\alpha r))^{(1+\delta)/2} P_0^{(\eta_1/2,\delta/2)}(\cos(2\alpha r)), \qquad (27a)$$

where $\eta_1 = \sqrt{9 + \frac{8mV_1}{\hbar^2\alpha^2}}$. For the illustration of this radial wave function, i.e., $R_{n,l}(r)/r$, of the trigonometric PT potential with various rotation-vibration $1s$, $1p$, $2s$, $2p$, $3s$ and $3p$ states, see the curves in Figure 5. Obviously, the number of nodes (in the allowed range) increases with the increasing of the vibration quantum number $n$. For example, the $1s$ and $1p$ states have one node, the $2s$ and $2p$ states have two nodes and so forth. That is, the wave functions of the rotating trigonometric Pöschl-Teller oscillator potential increase their oscillations with the increasing of the vibration quantum number $n$.

## 4. Final remarks and conclusion

In this work, we have obtained the approximate bound state solutions of the Schrödinger equation with the trigonometric Pöschl-Teller potential for arbitrary $l$-state in the framework of a new approximation for the centrifugal term $r^{-2}$. We employed a shortcut of the NU method in finding the energy eigenvalues and corresponding wave functions. Some numerical results are given in Table 1. It is found that in the limit when the potential range parameter $\alpha \to 0$, the energy levels approach to a constant value $\left(\sqrt{V_1}+\sqrt{V_2}\right)^2$. Under limiting case when $\alpha \to 0$, the trigonometric PT potential can be reduced to the Kratzer potential. We used a set of parameter values listed in Table 2 to calculate the energy spectrum of $I_2$, LiH, HCl, $O_2$, $H_2$, NO and CO diatomic molecules as illustrated in Table 3.

**Appendix A: Parametric Generalization of the NU method**

The NU method is used to solve second order differential equations with an appropriate coordinate transformation $s = s(r)$ [28]

$$\psi_n''(s) + \frac{\tilde{\tau}(s)}{\sigma(s)}\psi_n'(s) + \frac{\tilde{\sigma}(s)}{\sigma^2(s)}\psi_n(s) = 0, \tag{A1}$$

where $\sigma(s)$ and $\tilde{\sigma}(s)$ are polynomials, at most of second degree, and $\tilde{\tau}(s)$ is a first-degree polynomial. To make the application of the NU method simpler and direct without need to check the validity of solution. We present a shortcut for the method. So, at first we write the general form of the Schrödinger-like equation (B1) in a more general form applicable to any potential as follows [29]

$$\psi_n''(s) + \left(\frac{c_1 - c_2 s}{s(1 - c_3 s)}\right)\psi_n'(s) + \left(\frac{-As^2 + Bs - C}{s^2(1 - c_3 s)^2}\right)\psi_n(s) = 0, \tag{A2}$$

satisfying the wave functions

$$\psi_n(s) = \phi(s)y_n(s). \tag{A3}$$

Comparing (B2) with its counterpart (B1), we obtain the following identifications:

$$\tilde{\tau}(s) = c_1 - c_2 s, \quad \sigma(s) = s(1 - c_3 s), \quad \tilde{\sigma}(s) = -\xi_1 s^2 + \xi_2 s - \xi_3, \tag{A4}$$

Following the NU method [28], we obtain the following shortcut procedure [29]:

(i) The relevant constant:

$$c_4 = \frac{1}{2}(1 - c_1), \qquad c_5 = \frac{1}{2}(c_2 - 2c_3),$$

$$c_6 = c_5^2 + A, \qquad c_7 = 2c_4 c_5 - B,$$

$$c_8 = c_4^2 + C, \qquad c_9 = c_3(c_7 + c_3 c_8) + c_6,$$

$$c_{10} = c_1 + 2c_4 + 2\sqrt{c_8} - 1 \succ -1, \qquad c_{11} = 1 - c_1 - 2c_4 + \frac{2}{c_3}\sqrt{c_9} \succ -1, \; c_3 \neq 0,$$

$$c_{12} = c_4 + \sqrt{c_8} \succ 0, \qquad c_{13} = -c_4 + \frac{1}{c_3}(\sqrt{c_9} - c_5) \succ 0, \; c_3 \neq 0. \tag{A5}$$

(ii) The essential polynomial functions:

$$\pi(s) = c_4 + c_5 s - \left[\left(\sqrt{c_9} + c_3\sqrt{c_8}\right)s - \sqrt{c_8}\right], \quad (A6)$$

$$k = -(c_7 + 2c_3 c_8) - 2\sqrt{c_8 c_9}, \quad (A7)$$

$$\tau(s) = c_1 + 2c_4 - (c_2 - 2c_5)s - 2\left[\left(\sqrt{c_9} + c_3\sqrt{c_8}\right)s - \sqrt{c_8}\right], \quad (A8)$$

$$\tau'(s) = -2c_3 - 2\left(\sqrt{c_9} + c_3\sqrt{c_8}\right) \langle 0. \quad (A9)$$

(iii) The energy equation:

$$c_2 n - (2n+1)c_5 + (2n+1)\left(\sqrt{c_9} + c_3\sqrt{c_8}\right) + n(n-1)c_3 + c_7 + 2c_3 c_8 + 2\sqrt{c_8 c_9} = 0. \quad (A10)$$

(iv) The wave functions:

$$\rho(s) = s^{c_{10}}(1 - c_3 s)^{c_{11}}, \quad (A11)$$

$$\phi(s) = s^{c_{12}}(1 - c_3 s)^{c_{13}}, \quad c_{12} > 0, \ c_{13} > 0, \quad (A12)$$

$$y_n(s) = P_n^{(c_{10}, c_{11})}(1 - 2c_3 s), \quad c_{10} > -1, \ c_{11} > -1, \quad (A13)$$

$$\psi_{n\kappa}(s) = N_{n\kappa} s^{c_{12}}(1 - c_3 s)^{c_{13}} P_n^{(c_{10}, c_{11})}(1 - 2c_3 s). \quad (A14)$$

where $P_n^{(\mu,\nu)}(x)$, $\mu > -1$, $\nu > -1$, and $x \in [-1,1]$ are Jacobi polynomials with

$$P_n^{(\alpha,\beta)}(1 - 2s) = \frac{(\alpha+1)_n}{n!} {}_2F_1(-n, 1+\alpha+\beta+n; \alpha+1; s), \quad (A15)$$

and $N_{n\kappa}$ is a normalization constant. Also, the above wave functions can be expressed in terms of the hypergeometric function as

$$\psi_{n\kappa}(s) = N_{n\kappa} s^{c_{12}}(1 - c_3 s)^{c_{13}} {}_2F_1(-n, 1+c_{10}+c_{11}+n; c_{10}+1; c_3 s) \quad (A16)$$

where $c_{12} > 0$, $c_{13} > 0$ and $s \in [0, 1/c_3]$, $c_3 \neq 0$.

**Table 1** The bound state energy levels $E_{nl}$ for the trigonometric PT potential.

| state (n,l) | $E_{nl}$ | | | | | |
|---|---|---|---|---|---|---|
| | $M = 10 fm^{-1}, V_1 = 5 fm^{-1}, V_2 = 3 fm^{-1}$ [20] | | | | | |
| | $\alpha = 1.2$ | $\alpha = 0.8$ | $\alpha = 0.4$ | $\alpha = 0.2$ | $\alpha = 0.02$ | $\alpha = 0.002$ |
| | $d_0 = 0$ case | | | | | |
| 1s[21] | 22.87051710 | 20.32991862 | 17.95616357 | 16.83082621 | 15.85264289 | 15.75661628 |
| 2s[21] | 28.29143398 | 23.68420415 | 19.50420742 | 17.57271070 | 15.92394680 | 15.76371786 |
| 2p | 28.60696804 | 23.81346538 | 19.53372130 | 17.57973494 | 15.92401384 | 15.76371853 |
| 3s[21] | 34.28835086 | 27.2944896 | 21.11625126 | 18.33059518 | 15.99541071 | 15.77082105 |
| 3p | 34.63442822 | 27.43284957 | 21.14690619 | 18.33776218 | 15.99547790 | 15.77082171 |
| 3d | 35.31564933 | 27.70768323 | 21.20811765 | 18.35209065 | 15.99561226 | 15.77082304 |
| 4s[21] | 40.86126774 | 31.16077522 | 22.79229510 | 19.10447967 | 16.06703463 | 15.77792584 |
| 4p | 41.23788840 | 31.30823378 | 22.82409108 | 19.11178941 | 16.06710195 | 15.77792650 |
| 4d | 41.97876941 | 31.60107057 | 22.88757844 | 19.12640318 | 16.06723660 | 15.77792783 |
| 4f | 43.06137688 | 32.03529723 | 22.98255025 | 19.14830965 | 16.06743857 | 15.77792984 |
| $d_0 = 1/12$ case | | | | | | |
| 1s[21] | 22.87051710 | 20.32991862 | 17.95616357 | 16.83082621 | 15.85264289 | 15.75661628 |
| 2s[21] | 28.29143398 | 23.68420415 | 19.50420742 | 17.57271070 | 15.92394680 | 15.76371786 |
| 2p | 28.64395419 | 23.82847894 | 19.53712286 | 17.58054181 | 15.92402153 | 15.76371860 |
| 3s[21] | 34.28835086 | 27.2944896 | 21.11625126 | 18.33059518 | 15.99541071 | 15.77082105 |
| 3p | 34.67512504 | 27.44896381 | 21.15044543 | 18.33858626 | 15.99548560 | 15.77082179 |
| 3d | 35.43921159 | 27.75631556 | 21.21875330 | 18.35456399 | 15.99563534 | 15.77082328 |
| 4s[21] | 40.86126774 | 31.16077522 | 22.79229510 | 19.10447967 | 16.06703463 | 15.77792584 |
| 4p | 41.28229584 | 31.32544868 | 22.82776800 | 19.11263070 | 16.06710967 | 15.77792658 |
| 4d | 42.11348590 | 31.65300783 | 22.89862721 | 19.12892817 | 16.06725974 | 15.77792806 |
| 4f | 43.33519178 | 32.14003977 | 23.00470171 | 19.15336297 | 16.06748485 | 15.77793030 |

**Table 2** The spectroscopic parameters of diatomic molecules in the ground electronic state [21,22]. We take $V_2 = 0$.

| | $H_2$ | HCl | LiH | $I_2$ | $O_2$ | NO | CO |
|---|---|---|---|---|---|---|---|
| $V_1 = D_e$ (eV) | 4.744750871 | 4.619030905 | 2.515283695 | 1.581791863 | 5.156658828 | 8.043782568 | 10.84514471 |
| $\alpha^{-1} = r_e$ (Å) | 0.7416 | 1.2746 | 1.5956 | 2.662 | 1.208 | 1.1508 | 1.1282 |
| $\mu$ (amu) | 0.50391 | 0.98010 | 0.8801221 | 63.45223502 | 7.997457504 | 7.468441000 | 6.860586000 |

**Table 3** The rotation-vibration energy spectrum of several diatomic molecules in the ground electronic state.

| State | H$_2$ | HCl | LiH | I$_2$ | O$_2$ | NO | CO |
|---|---|---|---|---|---|---|---|
| 1s | 6.22078060 | 5.18052615 | 2.86602339 | 1.60083133 | 5.37162722 | 8.33502418 | 11.20478158 |
| 1p | 6.23937571 | 5.18330622 | 2.86801467 | 1.60084002 | 5.37199230 | 8.33545440 | 11.20526816 |
| 2s | 7.15241392 | 5.51563228 | 3.07657343 | 1.61176149 | 5.49641049 | 8.50374380 | 11.41288704 |
| 2p | 7.17235188 | 5.51850084 | 3.07863655 | 1.61177022 | 5.49677979 | 8.50417835 | 11.41337813 |
| 3s | 8.14902770 | 5.86123972 | 3.29458574 | 1.62272884 | 5.62262654 | 8.67415399 | 11.62290733 |
| 3p | 8.17030851 | 5.86419678 | 3.29672069 | 1.62273759 | 5.62300005 | 8.67459288 | 11.62340292 |

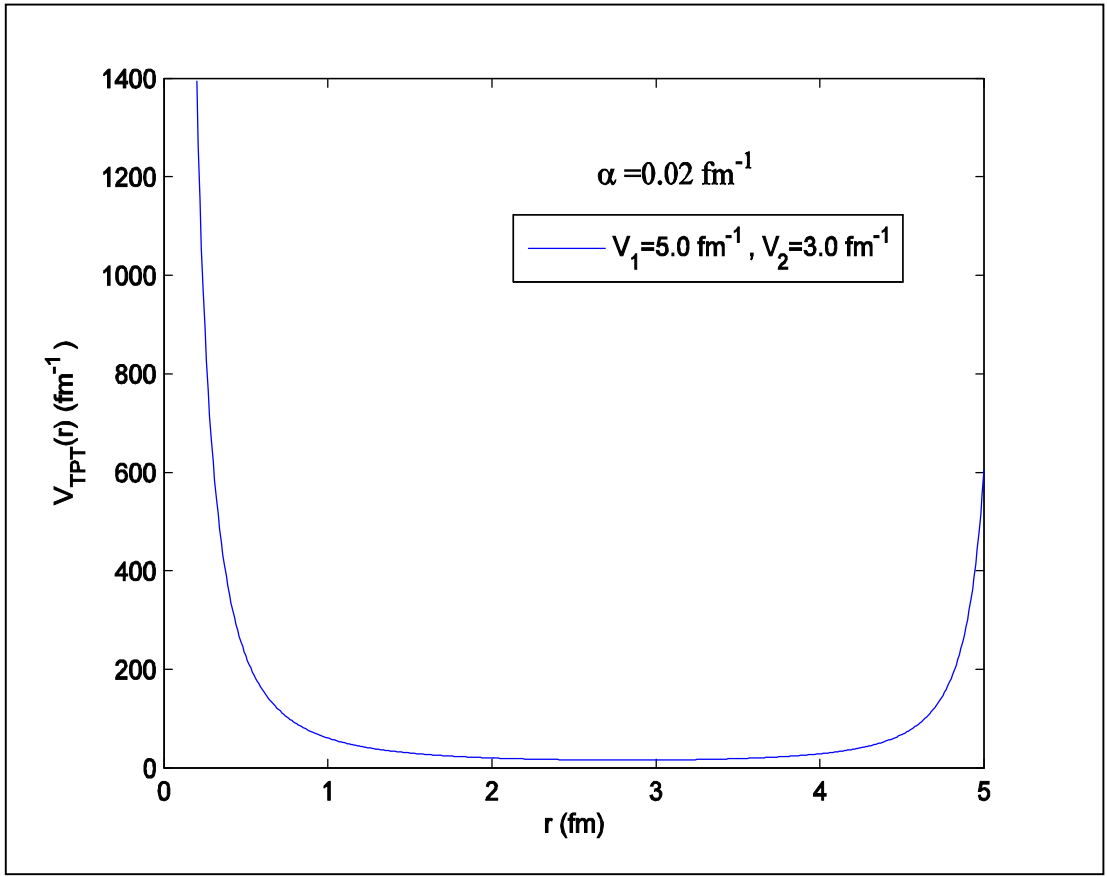

Figure 1. A draw of the trigonometric PT potential.

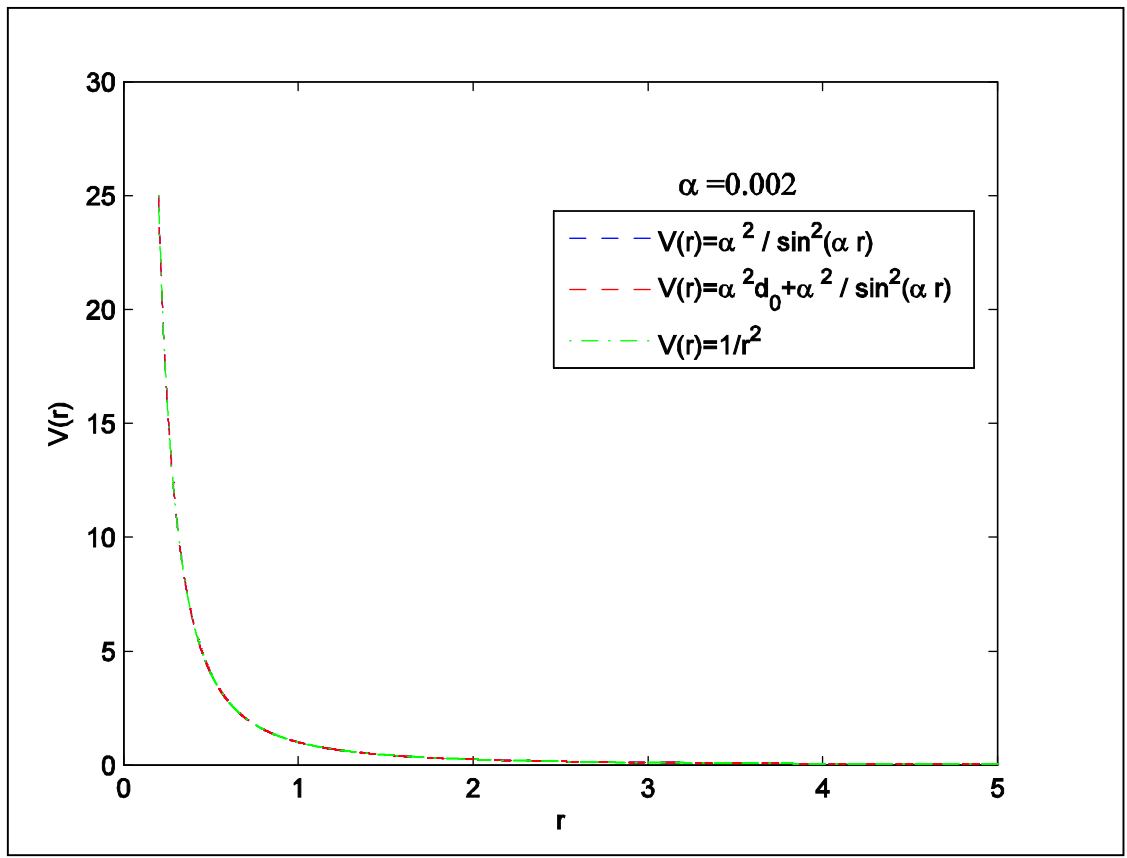

**Figure 2.** The centrifugal term $1/r^2$ (green line) and its approximations (14).

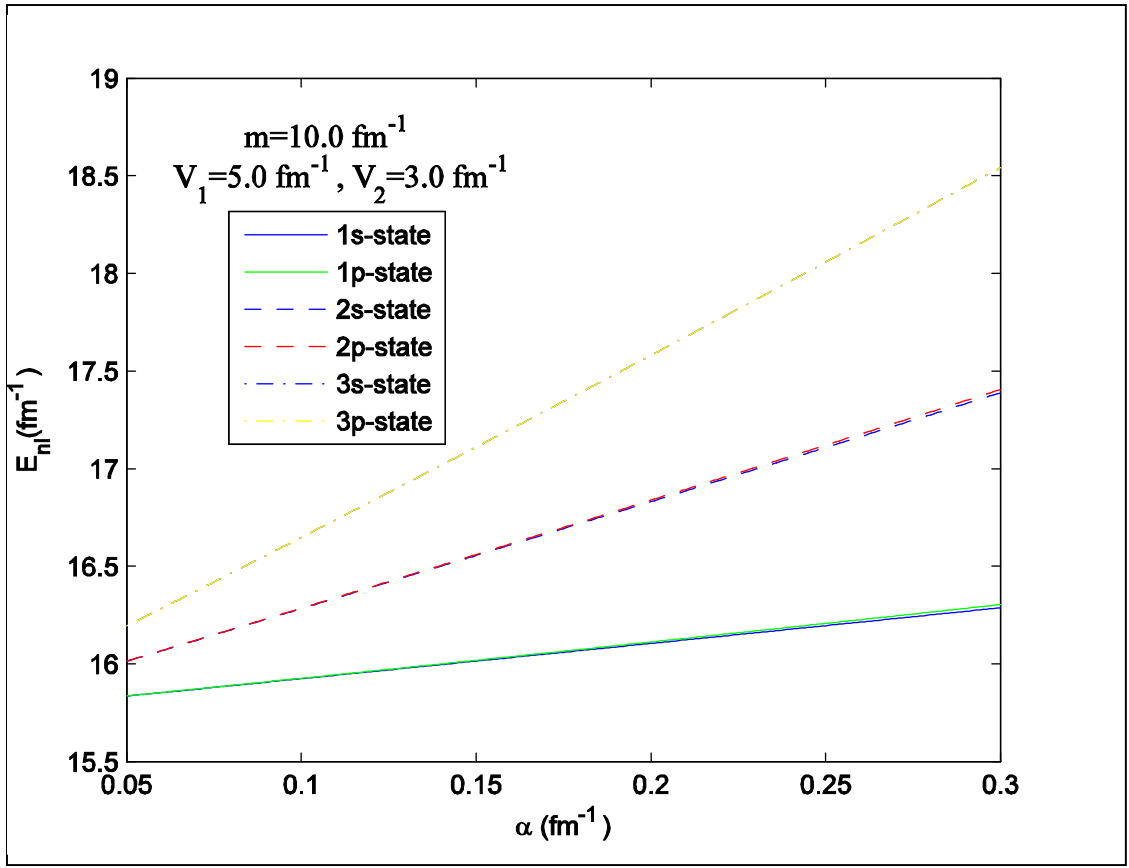

**Figure 3.** The variation of the lowest energy states $E_{nl}$ with the screening parameter $\alpha$. .

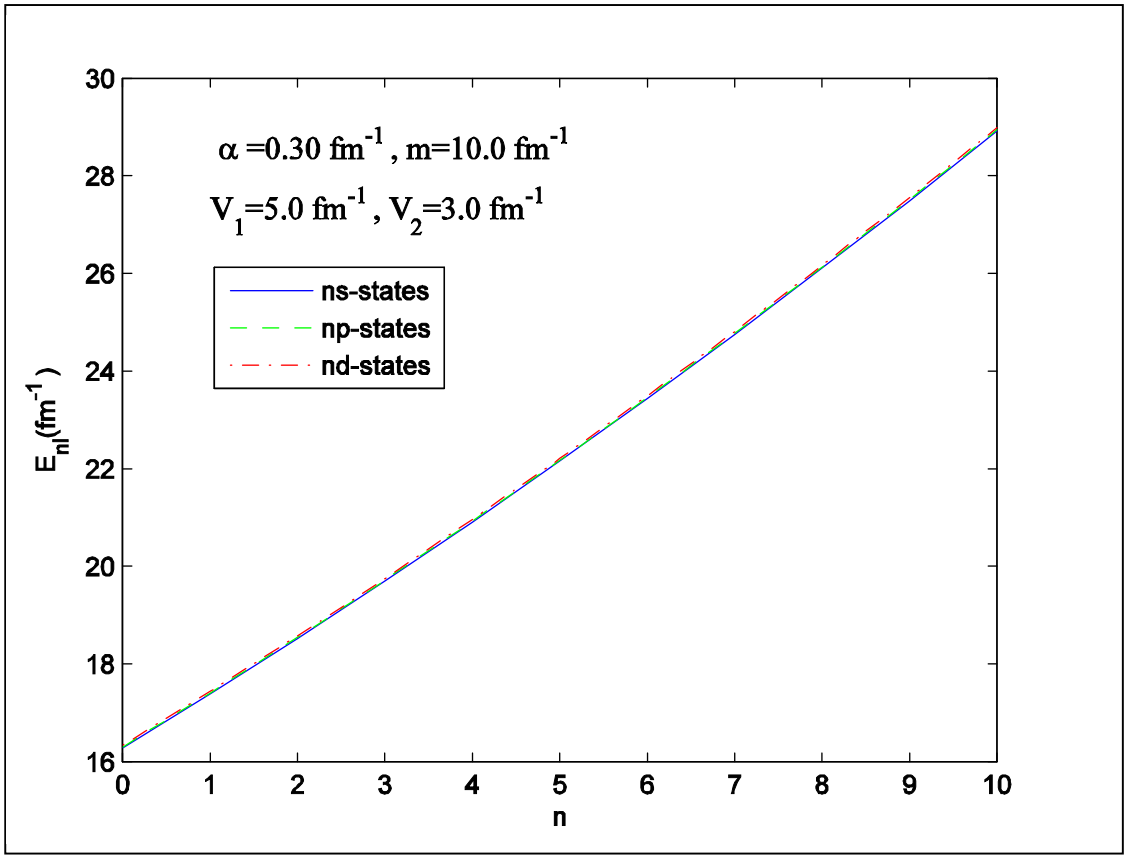

**Figure 4.** The variation of the lowest energy states $E_{nl}$ with the vibration quantum number $n$.

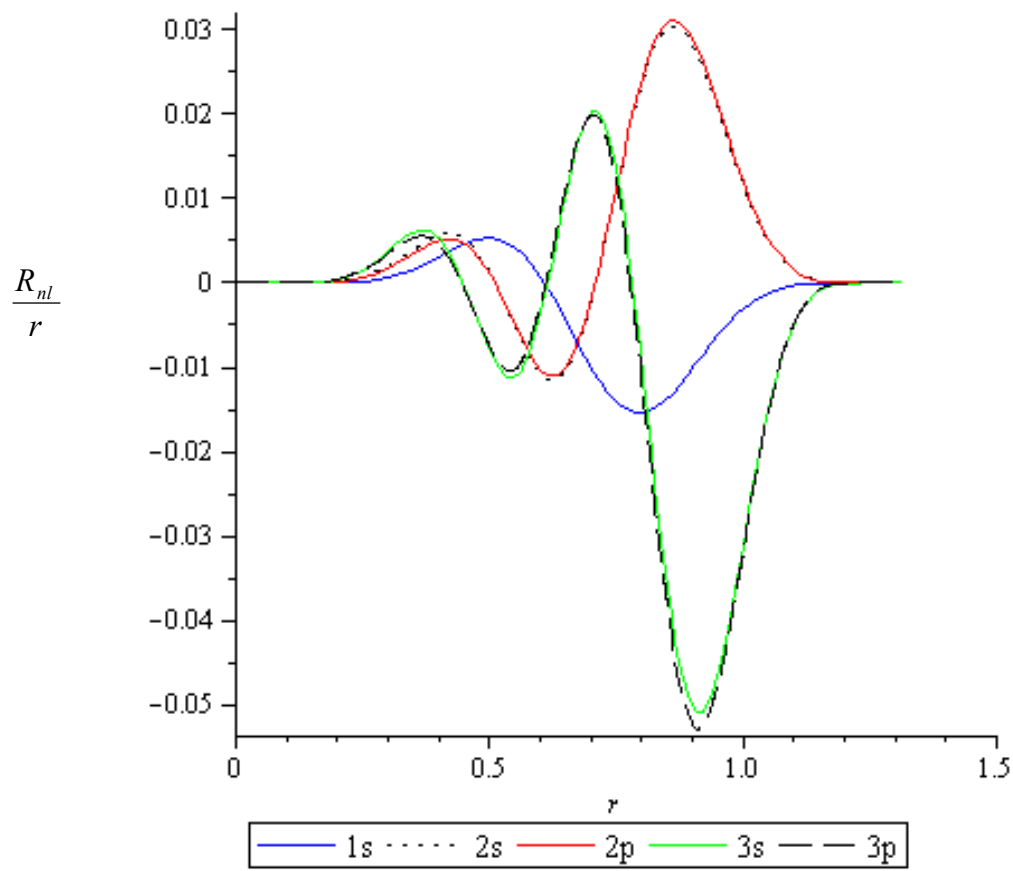

**Figure 5.** The radial wave functions of the trigonometric PT potential for the lowest vibration-rotation states.